\documentclass{ws-ijmpd}
\usepackage[super,compress]{cite}
\usepackage{graphicx}
\usepackage{amsmath}
\usepackage{amssymb}
\usepackage{amsxtra}

\def\({\left(}
\def\){\right)}

\def\beq{\begin{equation}}
\def\eeq{\end{equation}}
\begin{document}

\markboth{A. Chaudhuhri and M.Khlopov}
{Dark matter dilution scenarios in the early universe}

%
\catchline{}{}{}{}{}
%

\title{Dark matter dilution scenarios in the early universe}

\author{Arnab Chaudhuri}

\address{Department of Physics and Astronomy, Novosibirsk State University\\
Ul. Pirogova 2, Novosibirsk, 630090, Russia\\
arnabchaudhuri.7@gmail.com}

\author{Maxim Yu. Khlopov}

\address{National Research Nuclear University MEPhI (Moscow Engineering Physics Institute), 115409 Moscow, Russia\\APC laboratory 10,
rue Alice Domon et Leonie Duquet 75205 Paris Cedex 13, France\\
Institute of Physics,
Southern Federal University\\
Stachki 194
Rostov on Don 344090, Russia\\
khlopov@apc.univ-paris7.fr}

\maketitle

\begin{abstract}
When the vacuum like energy of the Higgs potential within the standard model
undergoes electroweak phase transition, an influx of entropy into the primordial
plasma can lead to a significant dilution of frozen out dark matter density
that was already present before the onset of the phase transition. The same
effect can take place, if the early Universe was dominated by
primordial black holes of small mass, evaporating before the period of Big Bang Nucleosynthesis. In this paper we
calculate the dilution factor for the above mentioned scenarios.
\end{abstract}


\section{Introduction}\label{intro}
The thermodynamic quantity, entropy, is a physical property of a system which can be roughly associated with the measure of disorder, chaos or randomness. This is yet another way to describe the second law of thermodynamics. By examining it, a direction of flow of energy or heat transfer, for example, hot to cold can be identified within a system. Following the example of the Carnot engine, i.e., for a reversible process, we have
\begin{equation} \label{carnot1}
\frac{Q_c}{Q_h}=\frac{T_c}{T_h},
\end{equation}
where $Q_c$ and $Q_h$ are the absolute values of heat transfer at temperatures $T_c$ and $T_h$ respectively. The ratio of $\frac{Q}{T}$ is defined as the change in entropy ($\Delta S$) for a reversible process and is given by:
\begin{equation} \label{entrp}
\Delta S=\left(\frac{Q}{T}\right)_{rev}.
\end{equation}
This result, which has general validity, means that the total change in entropy for a system in any reversible process is zero.

In thermal equilibrium with negligibly small chemical potential of the particles, the entropy conservation law follows \cite{1,2}, i.e.,
\begin{equation}
    S=\frac{\rho+\cal{P}}{T}a^3
\end{equation}
where $T(t)$ is the temperature of the plasma, $a(t)$ is the cosmological scale factor, $\rho$ and $P$ are the energy and pressure density respectively. Normally the state of matter is close to the equilibrium rate. In thermal equilibrium, the occupation number of any particle spices is given by two parameters, namely the chemical potential $\mu_j$ of each type of particles and the common temperature of all species. The Bose-Einstein condensate is an exception when chemical potential reaches the mass of the Boson particles, i.e., $\mu_j=m_B$. Even for this case the state of the system is determined by two parameters: the amplitude of the condensate and the temperature of the particles in the condensate. And for this case there is a rise in entropy $s$.

In thermal equilibrium, the distribution function is given by
\begin{equation}
    f(E)=\frac{1}{Exp(E/T) \pm 1}
\end{equation}
where $E$ is the particle energy.

The largest dilution of preexisting dark matter density due to entropy release can take during electroweak phase transition (EWPT). Theoretical calculations say that in the minimal standard model with one Higgs field the transition is the mild crossover \cite{3,4,5,6,7,8}.

There can be several realistic regimes during the universe history when this dilution occurred. One of them, for example, if the universe was at some epoch dominated by primordial black holes with small masses, the entropy production can be very high \cite{9}. In the presence of multi higgs field or a multi charged extension of the SM can lead to more rigorous results, \cite{A,B,C,D,E}.

Primordial black holes(PBH) might have dominated the cosmological energy density in the early universe and might have effected the baryon asymmetry of the universe, the fraction of dark matter particles, and would lead to the rise of the density perturbations at relatively small scales. Usually PBHs are created through Zel'dovich-Novikov (ZN) mechanism \cite{10,11}. According to ZN, a PBH can be created, if the density fluctuation $\delta \rho/\rho$ at the horizon size happened to be larger than unity.  With the accepted Harrison-Zel'dovich spectrum \cite{12,13} of primordial fluctuations the process of PBH creation can result in a significant density of PBHs. Generally the mass spectrum created by ZN mechanism is very narrow, approximated by the delta function. Other serious scenarios include the extended mass spectrum (Log-Normal distribution, etc.). The masses of PBH are assumed to be  sufficiently small, such that they evaporates sufficiently early, before the Big Bang Nucleosynthesis (BBN).

Though such short lived PBH would have decayed long before but their impact on the present day universe is noticeable. PBH decay can produce significant amount of entropy and hence dilute the baryon asymmetry. 

Initially the universe was dominated by relativistic particles/radiation (RD). The energy density is given by
\begin{equation}
    \rho_{rel}^{(1)}=\frac{3m_{pl}^2}{32\pi t^2}
\end{equation}
If sufficient amount of PBH were created and they survived till the moment when they became dominating in the universe, the cosmological expansion law turned into the non-relativistic one and the energy density
started to tend asymptotically to:
\begin{equation}
    \rho_{nr}=\frac{m_{pl}^2}{6\pi (t+t_1)^2}
\end{equation}
In usual baryogenesis, non-conservation of baryon numbers takes place at high temperatures whereas at low temperature it is conserved. 

In this work, the dilution of preexisting frozen out dark matter density for the above mentioned scenarios are dealt with. In the next section the theoretical framework of the work is given out followed by numerical calculation of the dilution factor. A generic conclusion follows. 

\section{Theoretical Framework}
\subsection{Dilution due to EWPT}
Consider the following Lagrangian:
\begin{equation} \label{1}
    L=\frac{1}{2}g^{\mu\nu}\partial_{\mu}\phi \partial_{\nu}\phi-U_{\phi}(\phi)+\sum_j i\left[g^{\mu\nu} \partial_{\mu}\chi_j^{\dagger}\partial_{\nu}\chi_j-U_j(\chi_j)\right]+L_{int}
\end{equation}
where $L_{int}$ is the Lagrangian of the Higgs boson with the field $\chi_j$ and is given by
\begin{equation} \label{2}
    L_{int}=\phi \sum_jg_j\chi_j^{\dagger}\chi_j
\end{equation}
and the summation is made all over relevant field $\chi_j$.

The self potential $\phi$ has the following form:
\begin{equation} \label{3}
    U(\phi)=\frac{\lambda}{4}(\phi^2-\eta^2)^2+\frac{T^2\phi^2}{2}\sum_jh_j\left(\frac{m_j(T)}{T}\right)
\end{equation}
where according to experiment the vacuum expectation value of $\phi$ is equal to $\eta=246$GeV and the quadratic self-coupling of $\phi$ is $\lambda=0.13$. Here $T$ is the plasma temperature and $m_j(T)$ is the mass of $\chi_j$-particle at $T$. The last term in eqn. \ref{3} appears as a result of thermal averaging of eqn. \ref{2}. The function $h_j(m_j/T)$ is positive and proportional to $T^2$. At high temperature it is multiplied by some constant but at low temperature it is exponentially suppressed. 

Most interest are given to the contributions from fermions. Their Yukawa coupling constant to Higgs are determined by their mass at zero temperature $m_f=g_f\eta$. The masses of the particles depends on the temperature because they are proportional to the vacuum expectation value and the later is proportional to the temperature dependent value of $\phi$ at the minima of the potential
\begin{equation} \label{4}
    \phi_{min}^2(T)=\eta^2-(T^2/\lambda)\sum_j h_j\left(\frac{m_j(T)}{T}\right)
\end{equation}
At temperature above than the critical temperature $T_c$, the expectation value of the Higgs field vanishes and below it, the expectation value gets a non zero value and the particles acquire mass. The transition temperature is given by 
\begin{equation} \label{5}
    T_c^2=\frac{\lambda \eta^2}{\sum h_j(0)}
\end{equation}
In terms of $T_c$, $\phi_{min}$ is given by
\begin{equation} \label{5}
    \phi_{min}^2=\eta^2\left[1-\frac{T^2}{T_c^2}\frac{\sum h_j(m_j/T)}{\sum h_j(0)}\right]=\eta^2\left[1-\frac{T^2}{T_c^2}\frac{h_{tot}(m/T)}{h_{tot}(0)}\right]
\end{equation}
To describe the behaviour of $h_{tot}(m)$, it is needed to define for each particles(fermion) the temperature at which the temperature dependent mass becomes equal to the  temperature, i.e., $m_f(T)=T$. Above this temperature, there are significant contributions from the particles. But below this temperature, the contributions are exponentially suppressed. 

The potential $U(\phi)$ is chosen in such a way that it vanishes when the field takes it's vacuum expectation value, i.e., $\phi=\eta$.It ensures zero vacuum energy of the classical field $\phi$. To calculate the entropy density, the energy momentum tensor is used:
\begin{eqnarray}
T_{\mu\nu} &=& \partial_\mu \phi \, \partial_\nu \phi - 
g_{\mu\nu} \left( g^{\alpha\beta} \partial_\alpha\phi \, \partial_\beta \phi - U_\phi (\phi) \right) 
\\ \nonumber 
&+& \sum_j \left[ \partial_\mu\chi_j^\dagger \, \partial_\nu\chi_j + \partial_\nu\chi_{j}^\dagger \, \partial_\mu \chi_{j} -
g_{\mu\nu} \left( g^{\alpha\beta} \partial_\alpha\chi^\dagger_{j} \, \partial_\beta\chi_{j} - U_j (\chi_j) + 2 {\cal L}_{int}  \right) \right] ,
\label{6}
\end{eqnarray}

The operators of the energy density and pressure density for homogeneous classical field $\phi$ and all other fields $\chi_j$ (quanta of $\phi$
should be  included there) in the Friedmann-Robertson-Walker background have the form:
\begin{eqnarray}
\rho &=& \dot\phi^2/2 + U_\phi (\phi)  + 
\sum_j \left[ \dot \chi_j^\dagger \dot \chi_j + \partial_l\chi_{j}^\dagger\, \partial_l\chi_{j }/a^2  + U_j (\chi_j) \right] - {\cal L}_{int}; \label{7} \\
{\cal P} &=& \dot\phi^2/2 - U_\phi (\phi)  + 
\sum_j \left[ \dot \chi_j^\dagger \dot \chi_j - (1/3)\partial_l\chi_{j}^\dagger\, \partial_l \chi_{j} /a^2  - U_j (\chi_j) \right] + {\cal L}_{int},
\label{8} 
\end{eqnarray}
where $\partial_i \chi$ is the space derivative. 

The sum of energy and pressure density enters into the expression for pressure density and is given by: 
\begin{equation}
    \rho + {\cal P} =  \dot\phi^2 + \sum_j \left[ \dot \chi_j^\dagger \dot \chi_j + 
\frac{2}{3 a^2}\, \partial_l\chi_{j}^\dagger \, \partial_l\chi_{j }   \right] .
\label{9}
\end{equation}

It can be verified that $\rho(t)$ satisfies the covarient conservation law:
\begin{equation} \label{9.5}
    \dot\rho = -3 H (\rho + {\cal P} )
\end{equation}

The calculations of the dilution factor as a result of entropy release  will be greatly simplified if we assume that the energy density
consists of two parts, the energy density of the field $\phi (t)$ sitting at the minimum of the potential 
and of relativistic matter, so the expression for $\rho$ becomes:
\begin{equation}
\rho \approx U_\phi (\phi_{min} ) + \frac{\dot \phi^2}{2}  + \frac{\pi^2 g_*}{30} T^4,
\label{10}
\end{equation}
and
\begin{equation}
\rho + {\cal P} \approx \dot \phi^2 + \frac{4}{3}\,\frac{\pi^2 g_*}{30} T^4,
\label{11}
\end{equation}
where $g_*\sim 10^2$ is the effective number of particle species at or near the electroweak phase transition.

The oscillations of $\phi$ around $\phi_{min} $ are quickly damped, so we take $\dot \phi = \dot \phi_{min}$ and neglect $\dot \phi^2$ 
in what follows, because the evolution of $\phi_{min}$ is induced by the universe expansion which is quite slow.
In this approximation we obtain the single differential equation governing the temperature evolution with time, or what is
more convenient, with the scale factor. Under this assumption eqn. \ref{9.5} can be written as:
\begin{equation} \label{12}
    \frac{\dot T}{T} \left[ 
h_{tot} (m) \eta^2  T^2 \left( 1  - \frac{T^2}{T^2_c}  \frac{ h_{tot} (m) }{ h_{tot}(0) } \right)  + \frac{4 \pi^2 g_*}{30} T^4\right]  
= - 4H \frac{\pi^2 g_*}{30} T^4 .
\end{equation}
It is to be noted that modification of the temperature due to annihilation of non-relativistic particles are ignored. Introducing the following parameters:
\begin{equation} \label{12.5}
    \kappa = \frac{30 h_{tot}(m)}{4 \pi^2 g_*}, \,\,\,\, \nu  = \frac{\kappa \eta ^2}{ 2 T_c^2}
\end{equation} 
And hence eqn. \ref{12} becomes
\begin{equation} \label{13}
\frac{\dot a }{a} = - \frac{\dot T}{T} \left[ \frac{\kappa \eta^2}{T^2} \left( 1 - \frac{T^2}{T_c^2}  \frac{ h_{tot} (m) }{ h_{tot}(0) } \right) 
 + 1 \right]
\end{equation}
In the case when the heaviest particle mass, that of $t$-quark, is lower than the temperature, we can take $h(m) = h (0)$ 
 and equation (\ref{13}) can be easily integrated resulting in:
 \begin{equation} \label{14}
     \frac{ a(T) T}{a_c T_c} = x^{2\nu} \exp\left[ \nu \left(\frac{1}{x^2} -1 \right)\right] 
 \end{equation}
 where $x = T/T_c$ and the cosmological scale factor $a_c$ is taken at $T=T_c$.
 
The largest contributing factor in this process is the heaviest particle, and in this case, $t$-quark. Its contribution to the entropy release is the largest
at high temperatures.

\subsection{Entropy production due to PBH decay}
Considering the simplest model of PBHs where they have a fixed mass $M_0$ with the number density at the moment of creation:
\begin{equation} \label{16}
    \frac{dN_{BH}}{dM} = \mu^3_1 \,\delta (M-M_0)
\end{equation}
where $\mu_1$ is a constant parameter with dimension of mass. All the PBHs were created at the same moment. Assume that the fraction of the 
PBH energy (mass) density at production was:
\begin{equation} \label{17}
    \frac{\rho_{BH}^{(in)}}{\rho_{rel}^{(in)}} = \epsilon \ll 1 
\end{equation}
Disregarding PBH decays, both relativistic particles and PBHs evolves independently.
\begin{equation} \label{18}
   \rho_{rel} (t)= \left(\frac{a^{(in)}}{a(t)}\right)^4 \rho_{rel}^{(in)},\,\,\,\,\,  
\rho_{BH} (t) = \left(\frac{a^{(in)}}{a(t)}\right)^3 \rho_{BH}^{(in)} 
\end{equation}
Let us consider the case when densities of relativistic and non-relativistic (PBH) matter became equal 
at $t = t_{eq}$ before the PBH decay. According to eqs.~(\ref{17}) and (\ref{18}) it takes place when:
\begin{equation}
\frac{\rho_{BH} (t_{eq})}{\rho_{rel} (t_{eq})} = \epsilon\, \frac{a(t_{eq})}{a_{in}} = 1 .
\label{19}
\end{equation}
It is assumed that at $t < t_{eq}$ the universe expansion is described by purely relativistic law and we find
\begin{equation} \label{20}
    t_{eq}=t_{in}/\epsilon^2
\end{equation}
PBHs would survive in the primeval plasma till equilibrium
if $t_{eq}- t_{in} < \tau_{BH}$, where the life-time of PBH with respect to evaporation is given by the expression:
\begin{equation}
\tau (M) \approx 3\times 10^3 N_{eff}^{-1} M_{BH}^3 m_{Pl}^{-4}  \equiv C\,\frac{M_{BH}^3}{m_{Pl}^4},
\label{21}
\end{equation}
where $N_{eff} \sim 100$ is the effective number of particle species with
masses smaller than the black hole temperature
\begin{equation}
T_{BH} = { m_{Pl}^2 \over 8\pi M_{BH}}
\label{22}
\end{equation}
Thus the condition that the RD/MD equality is reached prior to BH decay reads:
\begin{equation} \label{23}
    M_{BH} > \left[ \frac{m_{Pl}}{C} \left( \frac{1}{\epsilon^2} -1 \right)\right]^{1/2} \approx  \frac{m_{Pl} }{\sqrt{C} \,\epsilon }
\end{equation}
And after equilibrium the cosmological scale factor changes as:
\begin{equation} \label{24}
    a_{nr} (t) = a_{rel} (t_{eq}) \left(\frac{t+t_{eq}/3}{4t_{eq}/3}\right)^{2/3}
\end{equation}
The temperature of the relativistic plasma coexisting with the dominant PBH dropped down as the scale factor:
\begin{equation}
    T_{rel} = T_{eq} [a_{eq}/ a_{nr} (\tau) ] = T_{eq}\,\left(\frac{4 t_{eq}}{3\tau_{BH} + t_{eq}}\right)^{2/3} .
\label{25}
\end{equation}
Correspondingly the temperature of the newly created by the PBH decay relativistic plasma could be much higher than 
$T_{rel}$ given by eq.~(\ref{25}).
The suppression of dark matter density is directly related to the entropy suppression factor which is equal to the cube of the ratio of the temperatures of the "old"
relativistic plasma to that of the new one created by the PBH instant evaporation is: 
\begin{equation}
    \left( \frac{T_{rel}}{T_{heat}}\right)^3 = \left(\frac{a_{eq}}{a(\tau_{BH})}\right)^{3/4} = 
\left(\frac{4t_{eq}/3}{\tau_{BH} + t_{eq}/3}\right)^{1/2} = 
\frac{2}{\sqrt{3C}}\,\frac{m_{Pl}}{\epsilon M} \left( 1 + t_{eq}/3\tau \right)^{-1/2}.
\label{26}
\end{equation}

It's time to relax the instant decay approximation and solve numerically equations describing evolution of the cosmological energy densities
of non-relativistic PBHs and relativistic matter. It is convenient to work in terms of dimensionless time variable  $\eta = t/\tau_{BH}$, when
the equations take the form: 
\begin{equation}
  \frac{d \rho_{BH}}{d\eta} = - (3H\tau+ 1) \rho_{BH} 
  \label{27}
\end{equation}

\begin{equation}
    \frac{d \rho_{rel}}{d\eta} = - 4H\tau \rho_{rel} +  \rho_{BH}. 
    \label{28}
\end{equation}
We present the energy densities of PBH and relativistic matter  respectively in the form:
\begin{equation} \label{29}
    \rho_{BH} =\rho_{BH}^{(in)} \exp{(-\eta +\eta_{in})} y_{BH} (\eta)/ z(\eta)^{3}
\end{equation}
\begin{equation} \label{30}
    \rho_{rel} =\rho_{rel}^{(in)}  y_{rel} (\eta) /z(\eta)^{4}
\end{equation}
where $y_{rel}^{(in)} = y_{BH}^{(in)} = 1$ and 
\begin{equation}
    \eta_{in} = \frac{m_{Pl}^2}{C M^2_{BH} } \ll 1 .
\label{31}
\end{equation}
where C is defined above.

The redshift factor $z(\eta) = a(\eta)/a_{in}$ satisfies the equation:
\begin{equation}
    \frac{dz}{d\eta} = H \tau_{BH}\,z
\label{32}
\end{equation}
The Hubble parameter $H$ is determned by the usual expression for the spatially flat universe:
\begin{equation}
    \frac{3 H^2 m_{Pl}^2}{8\pi} = \rho_{rel} + \rho_{DM}.
\label{33}
\end{equation}
And the term $H\tau$ is expressed by the following expression:
\begin{equation}
    H \tau = \frac{C}{2}\,\frac{M_{BH}^2}{m_{Pl}^2}\,\left( \frac{y_{rel}}{z^4} + \frac{\epsilon}{ z^3 e^{\eta-\eta_{in}} } \right)^{1/2} .
\label{34}
\end{equation}
Evidently Eq. (\ref{27}) with $\rho_{BH} $ given by (\ref{29}) is solved as
\begin{equation}
    y_{BH}(\eta) = y_{BH}^{(in)} =1,
\label{35}
\end{equation}
while $\rho_{rel} (\eta) $ satisfies the equation:
\begin{equation}
    \frac{dy_{rel}}{d\eta} = \epsilon z (\eta) e^{-\eta + \eta_{in} }
\label{36}
\end{equation}
Equation \ref{32} and \ref{36} can be solved numerically with the initial conditions at $\eta = \eta_{in}$ 
\begin{equation}
    y_{bh}=y_{rel}=z=1 .
\label{37}
\end{equation}
However, a huge value of the coefficient  $H \tau$ makes the numerical procedure quite slow. To avoid that we introduce the new function 
$W$ according to:
\begin{equation}
    z = \sqrt{W}/\epsilon .
\label{38}
\end{equation}
and arrive to the equations:
\begin{eqnarray}
\frac{d W}{d\eta} &=& C \epsilon^2 \left(\frac{M}{m_{Pl}}\right)^2 \left( y_{rel} + {\sqrt{W}}\,e^{-\eta + \eta_{in} }\right)^{1/2} ,
\label{39}\\
\frac{d y_{rel}}{d\eta} &=& \sqrt{W} e^{-\eta + \eta_{in} },
\label{40}
\end{eqnarray}
where $W(\eta_{in}) =\epsilon^2 $. The diltion as a result of entropy release form PBH evaporation can be calculated as follows. In the absence of PBHs 
the quantities conserved in the comoving volume evolved as $1/z^3$. With extra radiation coming from the PBH evaporation
the entropy evolves as $y_{rel}^{3/4} /z^3 $. Hence the suppression of the relative number density of
frozen dark matter particles or earlier generated baryon asymmetry is equal to:
\begin{equation}
    S = \left[ y_{rel} (\eta ) \right]^{3/4}
    \label{41}
\end{equation}
when time tends to infinity. The temporal evolution of $S$ is depicted in the next section for different values of $\epsilon$ and $M_{BH}$.

\section{Calculations and Results}
\subsection{Calculation of dilution factor due to Electroweak phase transition}
We start the calculation of dilution factor with the calculations of the contribution of to the entropy from the heaviest particles.
To this end we need the numerical values of their coupling constants with the Higgs boson. According to 
the experimental data, they are 
$g_t^2 = 0.25$, $m_t^2 =173 $ GeV,
$\lambda = 0.13$, $g_W^ 2 = 0.13$, $g_Z =  0.1$. The Yukawa coupling constants of lighter 
fermions scale as the ratio of the masses, $g_f = g_t (m_f / m_t)$. 
With  the account of $t$-quark only the critical temperature is $T_c^2/\eta^2 =   2\lambda \eta^2/m_t^2  \approx 0.53$

The contribution of other heavy particle makes this ratio approximately twice lower:
\begin{equation}
    \frac{T_c^2}{\eta^2} =   \frac{ 2\lambda \eta^2}{m_t^2+m_w^2+m_z^2+m_H^2}  \approx .001 
\label{43}
\end{equation}
The next step is to estimate the values of $\kappa$ and $\nu$, see equation \ref{12.5}. As we mentioned before,
$h_{tot} (m)$ is a collection of theta-functions dominated by single contribution of a fermion. Correspondingly at $T> T^f_{min} \approx m_f$ the contribution of  a fermion to $h_{tot} (m)$ is equal to
$h_f = N_f  m^2 /(6\eta^2)$, where $N_q = 3$ and $N_l = 1$. Hence the contribution of all fermions lighter than $t$-quark is
\begin{equation}
    \kappa = \frac{5}{4\pi^2} \sum_f  \left[\frac{N_f  m^2_f }{ g_*( T_{min}^f ) \eta^2} \right],
\label{44}
\end{equation}
where  $g_*( T_{min}^f )$ is the number of relativistic particle species present in the plasma at $T \sim T_{min}^f$. So 
$\kappa$ and $\nu \approx \kappa$ are both small numbers, $\nu \approx 0.007$ for $g_* =106.76$ and 
$\nu \sim 0.07$ for $g_* = 10.75$. One should keep in mind, however, that these small numbers are multiplied by
$ (T_c /T_{min}^2)^2 \sim T_c^2/m_f^2 $, which can be very large. It is interesting that the product $\nu (T_c/m_f)^2$ 
essentially does not depend upon the mass and the effect is the larger for smaller masses due to a decrease of $g_*$.

The relative increase in entropy is given by the following expression:
\begin{equation}
    \frac{\delta s}{s} = \sum_f x_{j,min}^{6\nu_f} \exp\left[ 3\nu_f \left(\frac{1}{x_{f,min}^2} -1 \right)\right]  - 1 ,
\label{45}
\end{equation}
where $\nu_f$ includes only contribution from single fermion.

For example the electron contribution to the relative rise of the entropy is $(\delta s/s)_e = 1.8 \%$.  At  temperatures below the muon
mass $g_* =14.25$ and thus the muon contribution is $(\delta s/s)_\mu = 1.3 \%$. The contribution of $\tau$-lepton is 
$(\delta s/s)_\tau = 0.25 \%$, because at $T=180$ GeV $g_* = 75.75$. The contribution of four quark families 
in this temperature
range is 12 times larger and brings about $3\%$. The contribution from $t$-quark is $(\delta s/s)_t \approx 1\%$. The contribution from $b$-quark is almost the same but it is to be noted that the contribution came from a bigger range of temperature ($T_{min}$ for $b$-quark is $\approx 4 GeV$).  Moreover, the contributions of the  lighter $s$, $u$, and $d$ quarks are slightly 
enhanced because they remain alive down to the QCD phase transition at about 150 MeV, when $g_*$ is 72.25. The contribution of Higgs boson $(\delta s/s)_H = 1 \%$ and that of Gauge bosons is $(\delta s/s)_{W,Z} \approx 2 \%$.

Considering the contribution of entropy from the decoupled neutrinos at low temperature and other contributions at high temperature, we see the the net influx of entropy is $13\%$. Hence the the preexisting frozen out dark matter density has been diluted by $13\%$ by the influx of entropy into the primeval plasma as a result of EWPT withing the standard model.

\subsection{Entropy release due to PBH decay}
The suppression of the preexisting dark matter density by the influx of radiation/entropy created by the PBH decay as a function
of the PBH mass and $\epsilon = 10^{-12}$ is presented in the following figure \ref{1a}
\begin{figure}[h!]
\includegraphics[]{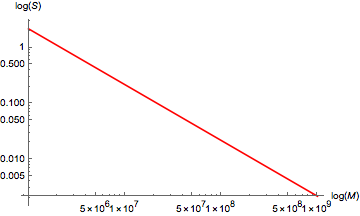}
\caption{Suppression factor of the cosmological dark matter density due to entropy influx from the PBH decay in the instant 
decay approximation.}
\label{1a}
\end{figure}
The PBH mass is bounded from below by the condition that the PBH did not decay before reaching equality with relativistic
matter. For $\epsilon = 10^{-12}$ it means that 
\begin{equation}
    M >   2 \cdot 10^6\,{\rm g} \left(10^{-12}/\epsilon \right) .
\label{46}
\end{equation}
The PBH mass is bounded from above by the condition that the heating temperature after evaporation should be higher 
than the BBN temperature, $\sim 1 $ MeV. Hence the PBH masses should be below $10^9$ g.  So the effect may be large for PBH masses in the range
$ 2\cdot 10^6 {\rm g} < M <   10^9 {\rm g} $.

Numerically solving equations \ref{39} and \ref{40}, the temporal evolution of S is depicted in the following figure \ref{1b} for $\epsilon=10^{-12}$ and for masses $10^8$g and $10^7$g. The red line line represents $10^8$g and the blue one represents $10^7$g. 
\begin{figure}[h!]
\includegraphics[]{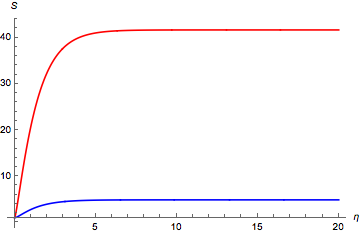}
\caption{Suppression factor of the cosmological dark matter density due to entropy influx from the PBH decay when instant decay approximation is relaxed.}
\label{1b}
\end{figure}
The results are in very good, about 15\%, agreement with the simplified estimates of the previous section.

\section{Conclusion}
We can see that for two different scenarios within the standard model there are possible ways to reduce the dark matter density, firstly by electroweak phase transition and secondly by the evaporation of primordial black holes. Even though EWPT in standard model is second order or smooth crossover, the entropy influx is significant. This dilution of frozen out species by the two mechanisms can lead to the dilution of preexisting baryon asymmetry which, might in turn, lead to Affleck-Dine baryogenesis and Lepto-through-baryo-genesis.

\section*{Acknowledgements}
The work of A.C. is funded by RSF Grant 19-42-02004. The research by M.K. was supported by the Ministry of Science and Higher Education of the Russian Federation under Project "Fundamental problems of cosmic rays and dark matter", No. 0723-2020-0040.


\end{document}